\documentclass[twocolumn,aps,prd,10pt,superscriptaddress,longbibliography,floatfix,citeautoscript]{revtex4-2}

\usepackage{amsmath}
\usepackage{amssymb}
\usepackage{bm}
\usepackage{dcolumn}
\usepackage{glossaries}
\usepackage{graphicx}
\usepackage{multirow}
\usepackage{fancyvrb}
\usepackage[linesnumbered,ruled]{algorithm2e}
\usepackage{siunitx}

\usepackage[usenames,dvipsnames,svgnames]{xcolor}
\usepackage{hyperref}
\hypersetup{
    pdfnewwindow=true,      
    colorlinks=true,        
    linkcolor=Blue,         
    citecolor=Blue,         
    filecolor=Blue,         
    urlcolor=Blue           
}

\usepackage{listings}	
\newacronym{ace}{ACE}{atomic cluster expansion}
\newacronym{dft}{DFT}{density functional theory}
\newacronym{emd}{EMD}{equilibrium molecular dynamics}
\newacronym{nemd}{NEMD}{non-equilibrium molecular dynamics}
\newacronym{hnemd}{HNEMD}{homogeneous non-equilibrium molecular dynamics}
\newacronym{md}{MD}{molecular dynamics}
\newacronym{mfp}{MFP}{mean-free path}
\newacronym{mlp}{MLP}{machine-learned potential}
\newacronym{nep}{NEP}{neuroevolution potential}
\newacronym{qhpf}{QHPF}{quasi-hexagonal-phase fullerene}
\newacronym{bpf}{BPF}{bulk-phase fullerene}
\newacronym{fcc}{FCC}{face-centered cubic}
\newacronym{rmse}{RMSE}{root mean square error}
\newacronym{snes}{SNES}{separable natural evolution strategy}
\newacronym{vdw}{vdW}{van-der-Waals}
\newacronym{lj}{LJ}{Lennard-Jones}

\DeclareSIUnit\angstrom{\text{Å}}
\DeclareSIUnit{\atom}{atom}
\DeclareSIUnit{\step}{step}
\DeclareSIUnit{\atomstepsecond}{\atom\step\per\second}

\newcolumntype{d}{D{.}{.}{-1}}

\begin{document}

\title{Anisotropic and high thermal conductivity in monolayer quasi-hexagonal fullerene: A comparative study against bulk phase fullerene}

\author{Haikuan Dong}
\affiliation{College of Physical Science and Technology, Bohai University, Jinzhou 121013, P. R. China}
\affiliation{Beijing Advanced Innovation Center for Materials Genome Engineering, Corrosion and Protection Center, University of Science and Technology Beijing, Beijing, 100083, China}

\author{Chenyang Cao}
\affiliation{Beijing Advanced Innovation Center for Materials Genome Engineering, Department of Physics, University of Science and Technology Beijing, Beijing 100083, China}

\author{Penghua Ying}
\email{hityingph@163.com}
\affiliation{School of Science, Harbin Institute of Technology, Shenzhen, 518055, P. R. China}

\author{Zheyong Fan}
\email{brucenju@gmail.com}
\affiliation{College of Physical Science and Technology, Bohai University, Jinzhou 121013, P. R. China}

\author{Ping Qian}
\email{qianping@ustb.edu.cn}
\affiliation{Beijing Advanced Innovation Center for Materials Genome Engineering, Department of Physics, University of Science and Technology Beijing, Beijing 100083, China}

\author{Yanjing Su}
\affiliation{Beijing Advanced Innovation Center for Materials Genome Engineering, Corrosion and Protection Center, University of Science and Technology Beijing, Beijing, 100083, China}

\date{\today}

\begin{abstract}
Recently a novel two-dimensional (2D) C$_{60}$ based crystal called quasi-hexagonal-phase fullerene (QHPF) has been fabricated and demonstrated to be a promising candidate for 2D electronic devices [Hou \textit{et al}. Nature \textbf{606}, 507-510 (2022)].
We construct an accurate and transferable machine-learned potential to study heat transport and related properties of this material, with a comparison to the face-centered-cubic bulk-phase fullerene (BPF).
Using the homogeneous nonequilibrium molecular dynamics and the related spectral decomposition methods, we show that the thermal conductivity in QHPF is anisotropic, which is $137(7)$ W/mK at 300 K in the direction parallel to the cycloaddition bonds and $102(3)$ W/mK in the  perpendicular in-plane direction. By contrast, the thermal conductivity in BPF is isotropic and is only $0.45(5)$ W/mK. We show that the inter-molecular covalent bonding in QHPF plays a crucial role in enhancing the thermal conductivity in QHPF as compared to that in BPF. The heat transport properties as characterized in this work will be useful for the application of QHPF as novel 2D electronic devices.
\end{abstract}
\maketitle


\section{Introduction}

Carbon has diverse chemical bonds and can form allotropes from three to zero dimension. A fullerene is a zero-dimensional allotrope consisting of carbon atoms connected by single and double bonds. The most typical fullerene C$_{60}$ has been extensively studied since its discovery \cite{kroto1985natur}. C$_{60}$ can form a single-crystal \cite{Tycko1991prl} with \gls{fcc} or simple-cubic structures, depending on the temperature. Thermal conductivity $\kappa$ has been used as a means to detect the ordering of the C$_{60}$ molecules in C$_{60}$ solids \cite{Yu1992prl}. It has been found that $\kappa \approx 0.4$ W/mK and is nearly temperature independent above 260 K. Above the critical temperature, the C$_{60}$ molecules begin to rotate quickly and phonons related to the center-of-mass translational degree of freedom are scattered by the rotational ones \cite{kumar2018prb}. With a few GPa compressing pressure, the C$_{60}$ fullerene system can be significantly hardened and $\kappa$ can be increased up to 5.5 W/mK at room temperature \cite{smontara1996thermal} and higher pressure can enhance $\kappa$ further \cite{giri2017prb}. Significant enhancement of $\kappa$ due to polymerization has also been predicted \cite{Abduljabar2018carbon}. These $\kappa$ values are much smaller than those in the quasi-one-dimensional and two-dimensional carbon allotropes, namely carbon nanotubes (CNTs) \cite{lee2017prl} and graphene \cite{balandin2008superior}. Adding functional groups can further reduce the thermal conductivity of fullerene-based materials \cite{Duda2013prl,Wang2013prb,chen2015thermal,Giri2017jpcl,Giri2020prm}. 

Despite the crystalline structures, heat transport in C$_{60}$ solids exhibit strong amorphous-like behaviors \cite{Giri2017jpcl,kumar2018prb}. In this paper, we show that a new form of C$_{60}$-based crystal has significantly larger $\kappa$ and exhibits crystalline behaviors for heat transport. This new crystal consists of a monolayer of C$_{60}$ molecules connected by covalent bonds, forming a structure named \gls{qhpf}. For convenience, we call the C$_{60}$-based bulk crystal as \gls{bpf}. \gls{qhpf} has been recently realized experimentally and shown to have good thermodynamic stability \cite{hou2022nature}. A transport bandgap of about 1.6 eV has been determined \cite{hou2022nature} and the in-plane structural anisotropy leads to anisotropic phonon modes and electrical conductivity. Theoretical calculations indicate that monolayer \gls{qhpf} is a promising candidate for photocatalysis \cite{peng2022monolayer}. However, the thermal transport properties of this novel material are still unknown. Due to the potential application of \gls{qhpf} in
2D electronic devices, it is important to characterize the thermal transport of this material.

Because the primitive cell of \gls{qhpf} contains 120 carbon atoms, \gls{md} is currently the only feasible computational approach \cite{gu2021jap} to theoretically study heat transport in this material in the diffusive transport regime. An important input to the \gls{md} approach is a classical interatomic potential. For carbon, there are a few important empirical potentials such as the Tersoff one \cite{tersoff1989prb,lindsay2010prb}, but they are mainly parameterized based on diamond and/or graphene structures, without being aware of the existence of the \gls{qhpf} structure. Recently, \glspl{mlp} have been shown to be a promising on-demand approach to achieve an accurate description of the potential energy surface of a general material, provided that a sufficiently large set of training structures with quantum-mechanical \gls{dft} data are available. 
Among the various \glspl{mlp}, the \gls{nep} approach \cite{fan2021neuroevolution,fan2022jpcm,fan2022arxiv} is one of the most computationally efficient. 

In this paper, we employ the \gls{nep} approach as implemented in the \textsc{gpumd} package \cite{fan2017cpc} to construct an accurate and transferable \gls{mlp} applicable to both \gls{qhpf} and \gls{bpf}, and study heat transport and related properties of \gls{qhpf}, with a comparison to \gls{bpf}. Using the \gls{hnemd} and the related spectral decomposition methods \cite{fan2019prb} , we show that the existence of the inter-molecular covalent bonds in \gls{qhpf} leads to significantly larger $\kappa$ in \gls{qhpf} as compared to \gls{bpf} in which the constituent C$_{60}$ molecules are mainly interacted by \gls{vdw} force. 

\begin{figure*}[t]
\centering
\includegraphics[width=1.6\columnwidth]{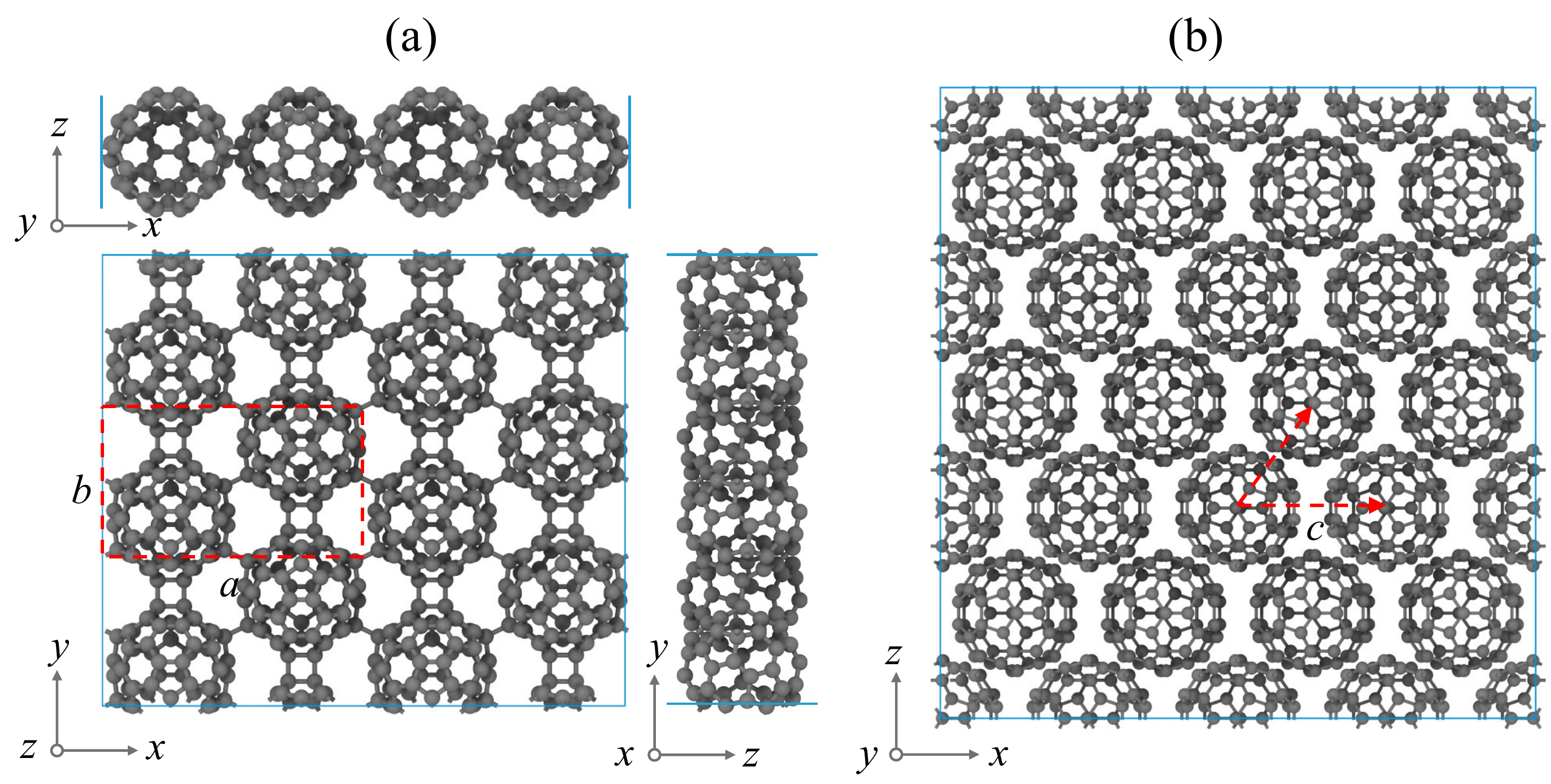}
\caption{
    Crystal structure of (a) monolayer \gls{qhpf} and (b) \gls{bpf}. The primitive cell of \gls{qhpf} is formed by the lattice parameters $a$ and $b$. 
}
\label{fig:model}
\end{figure*}

\section{Models and Methods}

\subsection{Models}

\subsubsection{The crystal structure of quasi-hexagonal-phase fullerene}

The crystal structure of monolayer \gls{qhpf} is schematically shown in Fig.~\ref{fig:model}(a). The dashed lines represent the primitive cell containing two C$_{60}$ molecules with $120$ carbon atoms in total. Each C$_{60}$ molecule is linked with six neighbouring ones by covalent bonds, with the $[2 + 2]$ cycloaddition of $'5, 6'$ bonds occurring along the $[010]$ direction and the C-C single-bonds forming along both the $[110]$ and $[1\overline{1}0]$ directions. Experimentally, 30-80 $\mu$m monolayer \gls{qhpf} sheets have been exfoliated \cite{hou2022nature}, demonstrating the thermal stability of this unique structure. 

\subsubsection{The crystal structure of bulk-phase fullerene}

We will comparably study the \gls{bpf} structure, which is schematically shown in Fig.~\ref{fig:model}(b). In this structure, each C$_{60}$ molecule occupies a site of the \gls{fcc} lattice. There is no clear covalent bonds between the C$_{60}$ molecules. Actually, the C$_{60}$ molecules do not have fixed orientations but rotate constantly at room temperature.

\subsection{The NEP approach for machine-learned potential}

The \gls{nep} approach for \gls{mlp} has been recently proposed \cite{fan2021neuroevolution} as a promising tool to study heat transport with high accuracy and low cost. It has been improved later \cite{fan2022jpcm,fan2022arxiv} and the version we used in this paper is the NEP3 model as detailed in Ref.~\cite{fan2022arxiv}.

For a \gls{mlp}, the descriptor \cite{Musil2021cr} is the most important aspect. In NEP3, the descriptor consists of a number of radial and angular components as described below. 

The radial descriptor components are constructed as 
\begin{equation}
\label{equation:qin}
q^i_{n}
= \sum_{j\neq i} g_{n}(r_{ij})
\quad\text{with}\quad
0\leq n\leq n_\mathrm{max}^\mathrm{R},
\end{equation}
where the summation runs over all the neighbors of atom $i$ within a certain cutoff distance.

For the angular descriptor components, we consider both 3-body ones ($0\leq n\leq n_\mathrm{max}^\mathrm{A}$, $1\leq l \leq l_\mathrm{max}^\mathrm{3b}$)
\begin{equation}
q^i_{nl} 
= \sum_{m=-l}^l (-1)^m A^i_{nlm} A^i_{nl(-m)},
\label{equation:qinl_spherical}
\end{equation}
and 4-body ones ($0\leq n\leq n_\mathrm{max}^\mathrm{A}$, $1\leq l_1=l_2=l_3 \leq l_\mathrm{max}^\mathrm{4b}$)
\begin{align}
q^i_{nl_1l_2l_3} 
=&
\sum_{m_1=-l_1}^{l_1}
\sum_{m_2=-l_2}^{l_2}
\sum_{m_3=-l_3}^{l_3}
\left(
\begin{array}{ccc}
l_1 & l_2 & l_3 \\
m_1 & m_2 & m_3\\
\end{array} 
\right) \nonumber \\
&\times A^{i}_{nl_1m_1}A^{i}_{nl_2m_2}A^{i}_{nl_3m_3}.
\label{equation:q_4body}
\end{align}
Here, 
\begin{equation}
A^i_{nlm} 
= \sum_{j\neq i} g_n(r_{ij}) Y_{lm}(\theta_{ij},\phi_{ij}),
\label{equation:Ainlm}
\end{equation}
and $Y_{lm}(\theta_{ij},\phi_{ij})$ are the spherical harmonics as a function of the polar angle $\theta_{ij}$ and the azimuthal angle $\phi_{ij}$ for the position difference $\bm{r}_{ij} \equiv \bm{r}_{j}-\bm{r}_{i}$ from atom $i$ to atom $j$. The 4-body descriptor components have been inspired by the \gls{ace} approach \cite{drautz2019prb}. 

The functions $g_n(r_{ij})$ in Eq.~(\ref{equation:qin}) are defined as a linear combination of $N_\mathrm{bas}^\mathrm{R}+1$ basis functions $\{f_k(r_{ij})\}_{k=0}^{N_\mathrm{bas}^\mathrm{R}}$:
\begin{align}
g_n(r_{ij}) &= \sum_{k=0}^{N_\mathrm{bas}^\mathrm{R}} c^{ij}_{nk} f_k(r_{ij}),\quad\text{with}
\label{equation:g_n}
\\
f_k(r_{ij}) &= \frac{1}{2}
\left[
    T_k\left(2\left(r_{ij}/r_\mathrm{c}^\mathrm{R}-1\right)^2-1\right)+1
\right]
f_\mathrm{c}(r_{ij}).
\label{equation:f_n}
\end{align}
Here, $T_k(x)$ is the $k^{\rm th}$ order Chebyshev polynomial of the first kind and $f_\mathrm{c}(r_{ij})$ is the cutoff function defined as
\begin{equation}
   f_\mathrm{c}(r_{ij}) 
   = \begin{cases}
   \frac{1}{2}\left[
   1 + \cos\left( \pi \frac{r_{ij}}{r_\mathrm{c}^\mathrm{R}} \right) 
   \right],& r_{ij}\leq r_\mathrm{c}^\mathrm{R}; \\
   0, & r_{ij} > r_\mathrm{c}^\mathrm{R}.
   \end{cases}
\end{equation}
Here, $r_\mathrm{c}^\mathrm{R}$ is the cutoff distance of the radial descriptor components.
The trainable expansion coefficients $c_{nk}^{ij}$ depend on $n$ and $k$ and also on the types of atoms $i$ and $j$. The functions $g_n(r_{ij})$ in Eq.~(\ref{equation:qinl_spherical}) and Eq.~(\ref{equation:q_4body}) are defined similarly but with a different basis size $N_{\rm bas}^{\rm A}$ and a different cutoff distance $r_{\rm c}^{\rm A}$.

The various descriptor components are grouped into a vector with $N_\mathrm{des}$ components,
$\{q^i_{\nu}\}_{\nu =1}^{N_\mathrm{des}}$. This vector is then taken as the input layer of a feedforward neural-network with a single hidden layer with $N_\mathrm{neu}$ neurons. The output of the neural network is taken as the potential energy of atom $i$. For the activation function in the hidden layer, we used the hyperbolic tangent function.

Although the neural network architecture is not different from the one as first proposed by Behler and Parrinello \cite{behler2007prl}, the neural network in \gls{nep} is trained using a novel evolutionary algorithm called 
the \gls{snes} \cite{Schaul2011}. The loss function guiding the training process is defined as a weighted sum of the \glspl{rmse} of energy, force, and virial as well as terms serving as $\ell_1$ and $\ell_2$ regularization. The weighting factors for these terms are denoted as $\lambda_\mathrm{e}$, $\lambda_\mathrm{f}$, $\lambda_\mathrm{v}$, $\lambda_1$, and $\lambda_2$, respectively. 

\subsection{The homogeneous nonequilibrium molecular dynamics method}

We used the \gls{hnemd} method to compute the thermal conductivity in a given system. This method was first formulated in terms of two-body potentials \cite{evans1982pla} and later generalized to many-body ones \cite{fan2019prb}, including \glspl{mlp} with atom-centered descriptors \cite{fan2021neuroevolution}. In this method, an external driving force (with zero net force) is added to the atoms in the system, leading to a heat current that has nonzero ensemble average (taken as time average in \gls{md} simulation) $\langle \bm{J} \rangle$. In the linear-response regime, the heat current is proportional to the driving force parameter $\bm{F}_{\rm e}$:
\begin{equation}
\label{equation:J}
\langle J^{\alpha} \rangle =  T V \sum_{\beta} \kappa^{\alpha\beta} F_{\rm e}^{\beta},
\end{equation}
where $T$ is the temperature and $V$ is the volume of the system. The proportionality constant $\kappa^{\alpha\beta}$ is the $\alpha\beta$ component of the thermal conductivity tensor. In this paper, we are only interested in diagonal principal components of the thermal conductivity tensor. Then in a given direction $\alpha$, the thermal conductivity component is written as $\kappa^{\alpha}\equiv \kappa^{\alpha\alpha}$ and is computed as
\begin{equation}
\label{equation:kappa}
\kappa^{\alpha} = \frac{\langle J_{\alpha}\rangle}{TVF_{\rm e}^{\alpha}}.
\end{equation}
The heat current can be resolved in the frequency domain, leading to the spectral thermal conductivity \cite{fan2019prb}
\begin{equation}
    \kappa^{\alpha}(\omega) = \frac{2}{VTF_{\rm e}^{\alpha}}\int_{-\infty}^{\infty} \text{d}t e^{\text{i}\omega t} \sum_i\sum_{j\neq i} \left\langle r^{\alpha}_{ij} \frac{\partial U_j}{\partial \bm{r}_{ji}}(0) \cdot \bm{v}_i(t) \right\rangle.
\end{equation}
Here, $U_j$ is the site energy of atom $j$, $\bm{r}_{ji}=\bm{r}_{i}-\bm{r}_{j}$, $\bm{r}_{i}$ is the position of atom $i$, and $\bm{v}_{i}$ is the velocity of atom $i$.

To cross check the \gls{hnemd} results, we also calculated the thermal conductivity of the \gls{qhpf} structure using the Green-Kubo relation in the \gls{emd} method: \cite{green1954jcp,kubo1957jpsj}
\begin{equation}
\kappa^{\alpha} = \frac{1}{k_{\rm B}T^2V} \int_0^{\infty} d\tau \langle J_{\alpha}J_{\alpha}(\tau)\rangle,
\end{equation}
where $k_{\rm B}$ is Boltzmann's constant and 
\begin{equation}
J_{\alpha} = \sum_i \sum_{j\neq i}  r^{\alpha}_{ij} \frac{\partial U_j}{\partial \bm{r}_{ji}} \cdot \bm{v}_i
\end{equation}
is the heat current that is applicable to general many-body potentials \cite{fan2015prb}.

In all the \gls{md} simulations, the time step for integration was set to 0.5 fs, the target temperature is 300 K and the target pressure is zero. In the \gls{hnemd} simulations the magnitude of the driving force parameter was chosen as $F_{\rm e}=0.5$ $\rm\mu m^{-1}$ and $F_{\rm e}=0.1$ $\rm\mu m^{-1}$  for \gls{bpf} and \gls{qhpf}, respectively. For \gls{bpf}, we have performed three  independent \gls{hnemd} simulations, each with a production time of 2 ns. For \gls{qhpf}, we have performed five independent \gls{hnemd} simulations, each with a production time of 5 ns. 
For the \gls{emd} method, we have performed 60 independent simulations, each with a production time of 5 ns.
The layer thickness of \gls{qhpf} was set to 8.785 \AA{}, which is the distance between two layers in bulk \gls{qhpf} \cite{hou2022nature}.

\begin{figure*}
\centering
\includegraphics[width=2\columnwidth]{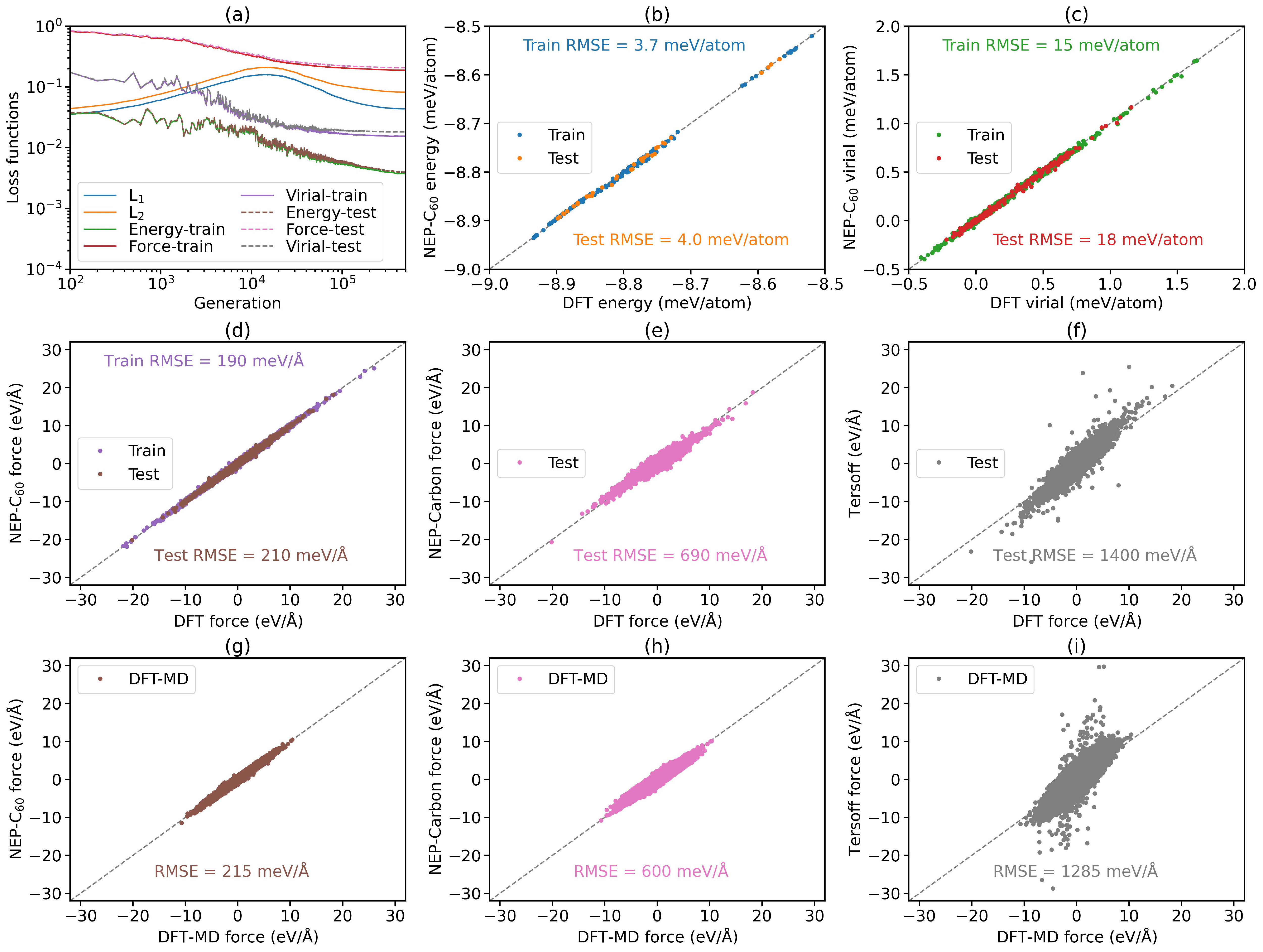}
\caption{
    (a) Evolution of the various terms in the loss function for the training and testing data sets with respect to the generation. (b) Energy, (c) virial, and (d) force calculated from NEP-C$_{60}$ as compared to the DFT reference data for the training and testing data sets. Force calculated from (e) NEP-Carbon and (f) Tersoff as compared to the DFT reference data for the testing data set. Force calculated from (g) NEP-C$_{60}$, (h) NEP-Carbon and (i) Tersoff as compared to the DFT-MD reference values in the extra testing data set (1000 structures).
}
\label{fig:train}
\end{figure*}

\begin{figure}
\centering
\includegraphics[width=0.8\columnwidth]{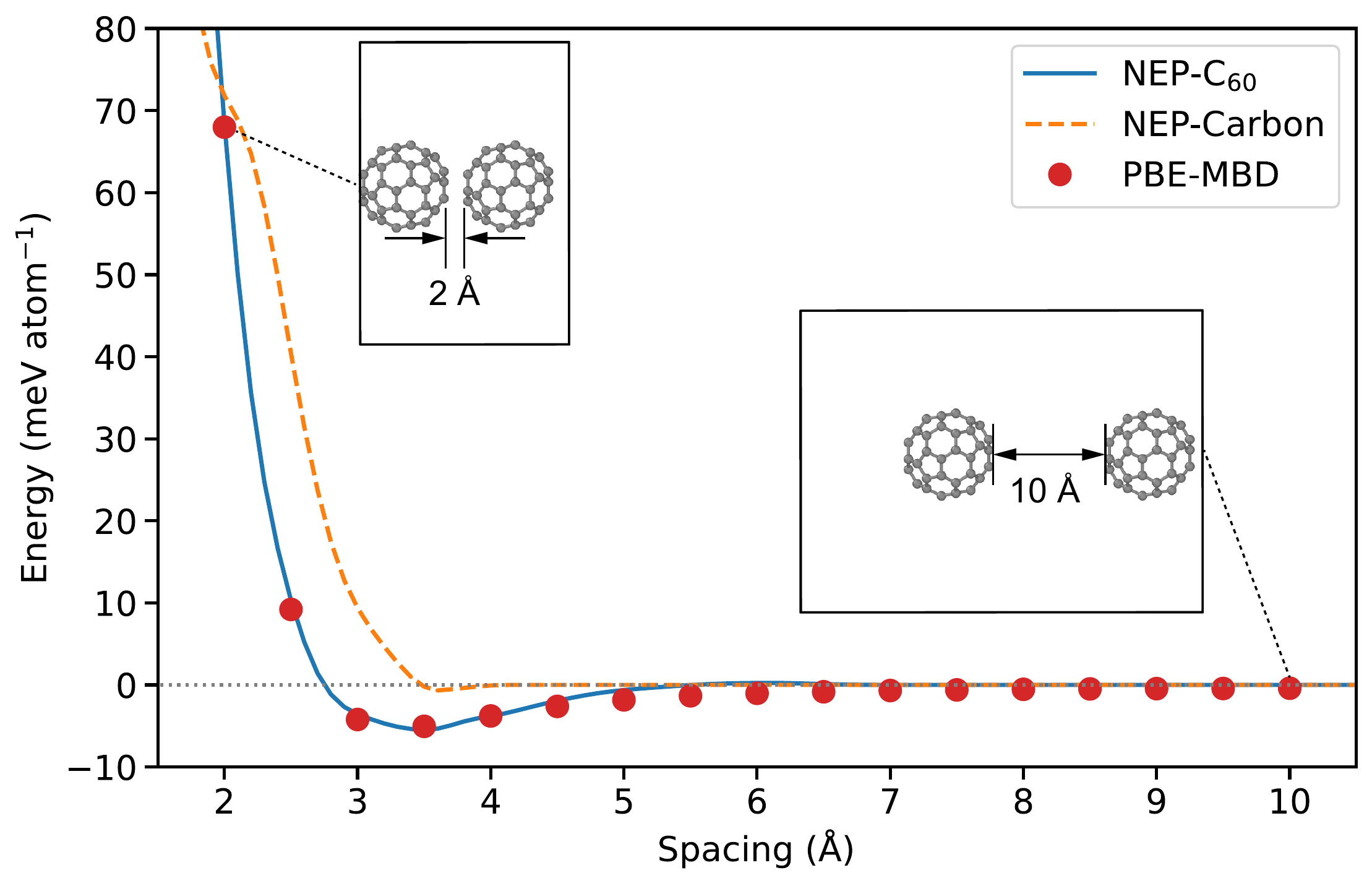}
\caption{
    Energy of a linear chain of C$_{60}$ molecules as a function of the inter-molecular spacing.
}
\label{fig:dimer}
\end{figure}

\section{Results and Discussion}

\subsection{Training and validating the new NEP model}

A \gls{nep} model for carbon systems has been recently trained \cite{fan2022arxiv}, but the training data were mostly consisting of diamond, graphite, amorphous, and liquid structures, without explicit C$_{60}$ structures \cite{Deringer2017prb}. Both the \gls{qhpf} and the \gls{bpf} structures are stable in \gls{md} simulations with this \gls{nep} model, but as we will show later, this \gls{nep} model does not have sufficiently high accuracy for the \gls{qhpf} and \gls{bpf} structures. To ensure high accuracy in the \gls{md} simulations, we develop a more accurate \gls{nep} model trained against \gls{qhpf} and \gls{bpf} structures with \gls{dft} reference values. For clarity, we call the old and new \gls{nep} models \gls{nep}-Carbon and \gls{nep}-C$_{60}$, respectively.

\subsubsection{Generation of training and testing structures}

The training and testing structures include \gls{qhpf} structures, \gls{bpf} structures, and C$_{60}$ chains. 

For \gls{qhpf}, we used the \gls{nep}-Carbon model \cite{fan2022arxiv} to run $NpT$ (constant number of atoms $N$, controlled pressure $p$, and controlled temperature $T$) simulations with a rectangular box containing 120 carbon atoms. The Bussi-Donadio-Parrinello thermostat \cite{Bussi2007jcp} and the Bernetti-Bussi barostat \cite{Bernetti2020jcp} were used to realize the $NpT$ ensemble. We considered three target pressures: 0, 1, and $-1$ GPa in both the $x$ and $y$ directions. For each target pressure, we linearly increased the target temperature from 10 K to 1000 K during a simulation time of 2,500 ps. We sampled the structures every 50 ps, obtaining 50 structures for each target pressure. Therefore, we have collected 150 structures in total. We used 120 randomly selected structures for training and the remaining 30 for testing. Apart from the above, we also performed \gls{dft}-based \gls{md} simulations at 800 K for 10 ps and sampled 1000 structures (each containing 240 atoms) to be used as an extra testing data set.

For \gls{bpf}, we started from an ideal FCC unit cell with four C$_{60}$ molecules, which contains 240 carbon atoms in total. The centers of two neighboring C$_{60}$ molecules are separated by 9 \AA{} \cite{giri2017prb}. We then applied perturbations with $3\%$ random box deformations and $0.1$ \AA{} random atom displacements to create 15 training structures and 4 testing ones. 

For C$_{60}$ chains, we considered two C$_{60}$ molecules in a box with vacuum in the transverse directions and varied the separation between them from 2 to 10 \AA{}, with an interval of 0.5 \AA, obtaining 17 structures in total. All these structures were used for training. 

In summary, we have 152 and 1034 structures in the training and testing data sets, which contain 20,040 and 244,560 atoms, respectively. We have checked the learning curve to confirm that this training data set is large enough.

\subsubsection{DFT calculations}

After obtaining the structures, we used quantum-mechanical \gls{dft} calculations to obtain their reference energy, force, and virial data. To this end, we used the \textsc{vasp} package \cite{Kresse1996prb} and the PBE functional \cite{Perdew1996prl} combined with the many-body dispersion correction \cite{Tkatchenko2012prl}. The energy cutoff for the projector augmented wave \cite{paw1, paw2} was chosen as 650 eV. A $\Gamma$-centered $k$-point mesh with a density of $0.25$ \AA$^{-1}$, and a threshold of $10^{-8}$ eV were used for the electronic self-consistent loop. A Gaussian smearing with a width of 0.1 eV was used. 

\subsubsection{Choosing the training hyperparameters}

After calculating the reference values, we used the \textsc{gpumd} package \cite{fan2017cpc,fan2022arxiv} to train the \gls{nep}-C$_{60}$ model. The hyperparameters we chose are listed in Table \ref{table:hyper}. Compared to the hyperparameter values for the \gls{nep}-Carbon model \cite{fan2022arxiv}, we have the following modifications. First, we have increased the radial cutoff from 4.2 to 7 \AA{} and increased the angular cutoff from 3.7 to 4 \AA{}. The notable increase in the radial cutoff is justified by the need for accurately describing the \gls{vdw} interactions between the C$_{60}$ molecules, as will be further discussed below. Second, we have removed the 5-body descriptor components as defined in Ref.~\cite{fan2022arxiv}, which are not very important for our system. Third, we have increased the $\ell_1$ and $\ell_2$ regularization weights from 0.05 to 0.1, which can help to increase the robustness of the potential in \gls{md} simulations. Fourth, we have increased the virial weight in the loss function from 0.1 to 0.5, because virial has been regarded to be important for heat transport applications \cite{shimamura2020jcp}.

\begin{table}[htb]
\centering
\setlength{\tabcolsep}{2.5Mm}
\caption{Hyperparameters for the \gls{nep}-C$_{60}$ model.}
\label{table:hyper}
\begin{tabular}{lllllll}
\hline
\hline
parameter & value & parameter & value\\
\hline
$r_{\rm c}^{\rm R}$ & 7 \AA & $r_{\rm c}^{\rm A}$ & 4 \AA \\
$n_{\rm max}^{\rm R}$ & 10 & $n_{\rm max}^{\rm A}$ & 8 \\
$N_{\rm bas}^{\rm R}$ & 10 & $N_{\rm bas}^{\rm A}$ & 8 \\
$l_{\rm max}^{\rm 3b}$ & 4 & $l_{\rm max}^{\rm 4b}$ & 2 \\
$N_{\rm neu}$ & 50 & $\lambda_{1}$ & 0.1 \\
$\lambda_{2}$ & 0.1 & $\lambda_{\rm e}$ & 1.0 \\
$\lambda_{\rm f}$ & 1.0 & $\lambda_{\rm v}$ & 0.5 \\
$N_{\rm bat}$ & 10000 & $N_{\rm pop}$ & 60 \\
$N_{\rm gen}$ & $5 \times 10^5$ \\
\hline
\hline
\end{tabular}
\end{table}

\subsubsection{Training and testing results}

Figure \ref{fig:train}(a) shows the evolution of the various components in the loss function during the training process. The training has been performed for $5\times 10^5$ generations, after which the predicted energy, virial, and force are compared against the \gls{dft} reference values in Figs.~\ref{fig:train}(b)-\ref{fig:train}(d) and \ref{fig:train}(g). The \glspl{rmse} of the various quantities for both the training data set and the hold-out testing data set are presented. As a comparison, we also show the parity plots of force for the NEP-Carbon model and the Tersoff potential in Figs.~\ref{fig:train}(e), \ref{fig:train}(f), \ref{fig:train}(h), and \ref{fig:train}(i). We see that the current NEP model exhibits a much higher accuracy compared to the other two potential models. Our \gls{nep}-C$_{60}$ model can also accurately describe the interaction energy between the C$_{60}$ molecules in the linear chain structure with different inter-molecular distances, as shown in Fig.~\ref{fig:dimer}. The \gls{vdw} interactions between the C$_{60}$ molecules are mostly captured by the radial descriptor components with a relatively long  cutoff (7 \AA{}) in our \gls{nep} model. In other \glspl{mlp}, \gls{vdw} interactions in carbon systems have been modelled by an explicit dispersion term with \cite{Muhli2021prb} or without \cite{wen2019prb,Rowe2020jcp} environment dependence. 

\begin{table}[htb]
\centering
\setlength{\tabcolsep}{2.5Mm}
\caption{Lattice constants of monolayer \gls{qhpf} as calculated by the various potentials as well as \gls{dft}.}
\label{table:lattice_compare}
\begin{tabular}{lllllll}
\hline
\hline
Model & $a$ (\AA) & $b$ (\AA)  \\
\hline
\gls{dft} & 15.81 & 9.12  \\
\hline
\gls{nep}-C$_{60}$ & 15.79 & 9.14  \\
\hline
\gls{nep}-Carbon & 15.88 & 9.18  \\
\hline
Tersoff & 16.30 & 9.35  \\
\hline
\hline
\end{tabular}
\end{table}

\begin{figure}
\centering
\includegraphics[width=\columnwidth]{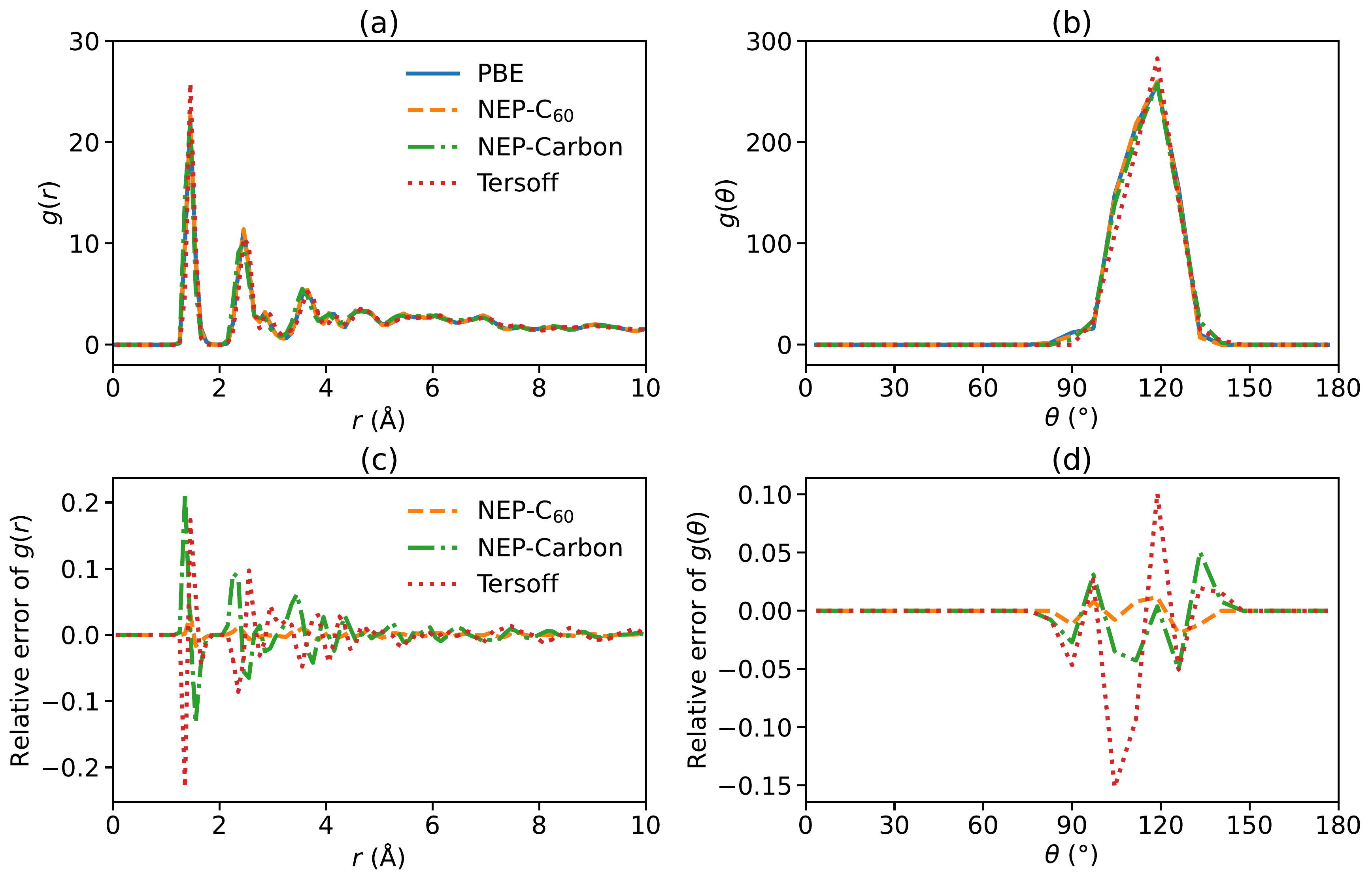}
\caption{
The (a) radial and (b) angular distribution functions of the \gls{qhpf} structure at 800 K from DFT-PBE and the three potentials. The relative errors of the predictions by the three potentials to the DFT-PBE results for the (c) radial and (d) angular distribution functions.}
\label{fig:rdf}
\end{figure}

\begin{figure}
\centering
\includegraphics[width=\columnwidth]{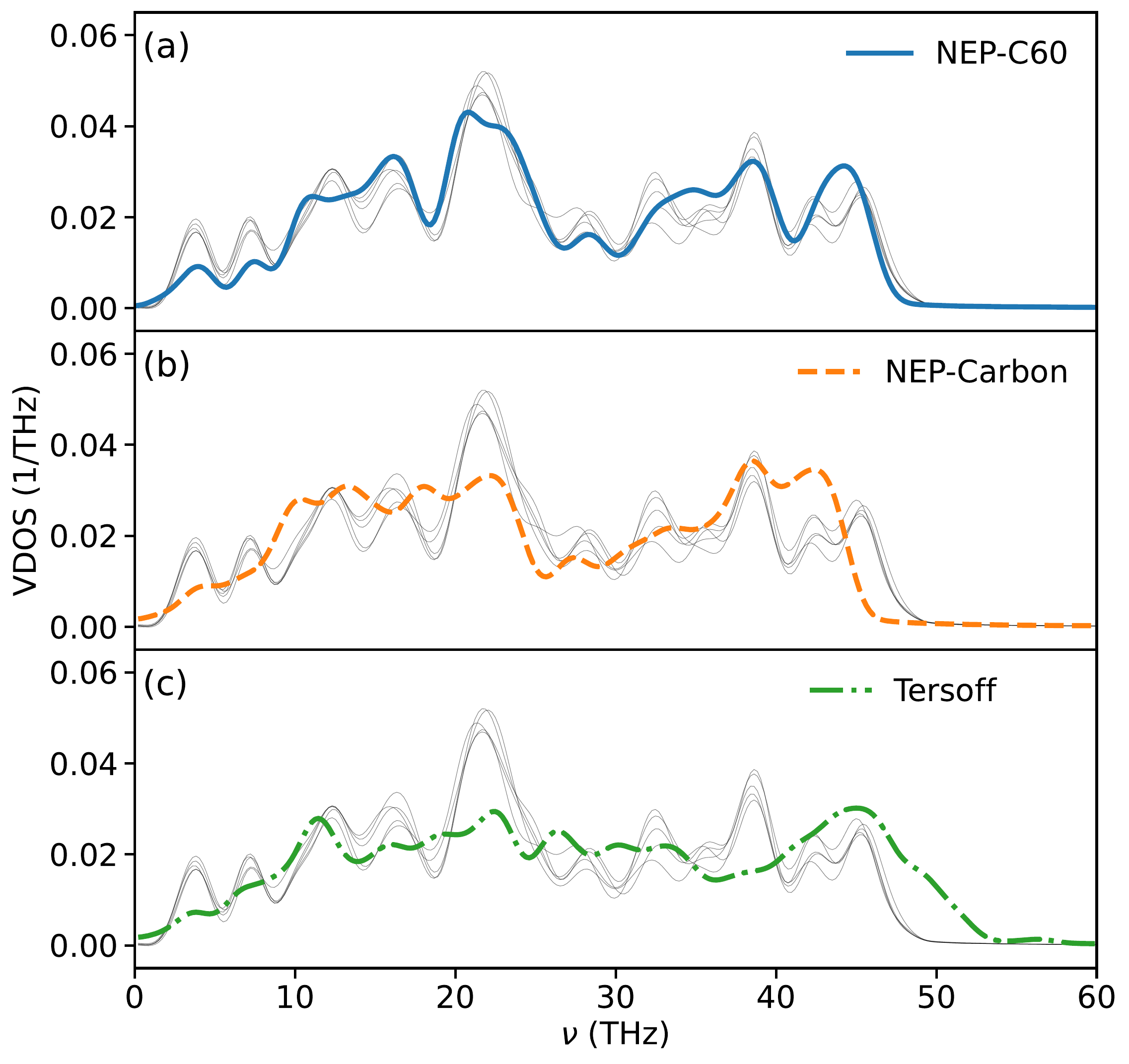}
\caption{
The vibrational density of states of the \gls{qhpf} structure at 800 K calculated using the (a) NEP-C$_{60}$ model, (b) the NEP-Carbon model, and (c) the Tersoff potential. Due to the high computational cost for DFT-PBE, we show the results from five short trajectories (represented as five thin lines in each subplot) to emphasize its relatively large statistical errors. For the three classical potential models, we have used long trajectories and the results have negligible statistical errors which are thus not shown for clarity.
}
\label{fig:vdos}
\end{figure}


To further demonstrate the higher accuracy of the \gls{nep}-C$_{60}$ model compared to the \gls{nep}-Carbon model and the Tersoff potential, we compare the lattice constants in Table \ref{table:lattice_compare}. The lattice constants calculated from the \gls{nep}-C$_{60}$ model agree with the \gls{dft} values with less than $0.1\%$ errors. The \gls{nep}-Carbon model has a little larger but still acceptable errors, but the Tersoff potential has 2-3 \% errors. Figures \ref{fig:rdf} and \ref{fig:vdos} show that the NEP-C$_{60}$ model also has higher accuracy than the other two for radial distribution, angular distribution, and vibrational density of states. We expect that the higher accuracy of the \gls{nep}-C$_{60}$ model can lead to more reliable prediction of the physical properties as discussed in the remainder of this paper.

\subsection{Thermal transport}

\subsubsection{The bulk-phase fullerene crystal}

\begin{figure}
\centering
\includegraphics[width=\columnwidth]{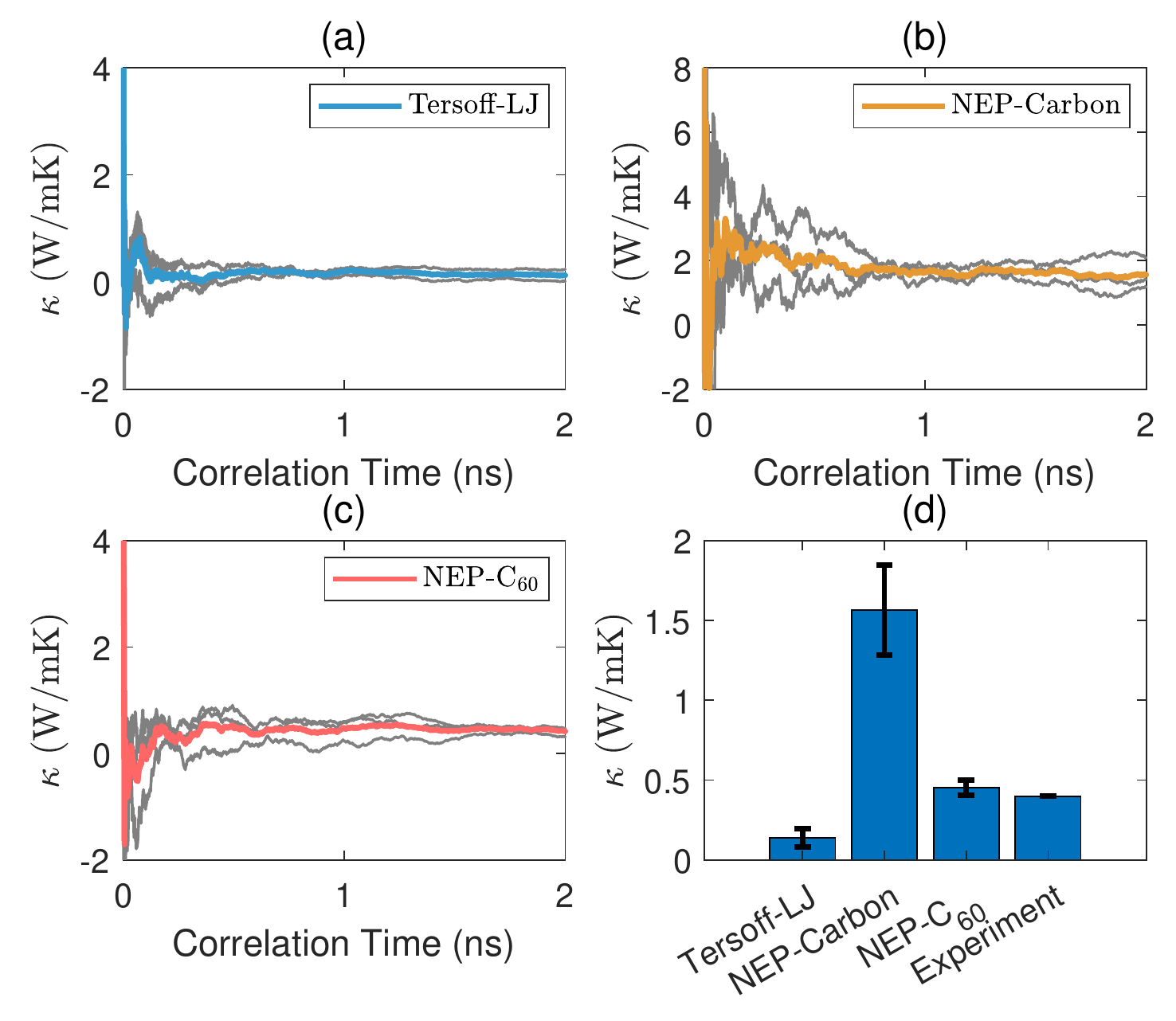}
\caption{
Thermal conductivity as a function of time for BPF at 300 K calculated using the \gls{hnemd} method with (a) the Tersoff-\gls{lj} potential, (b) the \gls{nep}-Carbon model, and (c) the \gls{nep}-C$_{60}$ model. (d) The converged thermal conductivity values from the different potentials as compared to an experiment value \cite{Yu1992prl}. 
}
\label{fig:fcc_kappa}
\end{figure}

Figures~\ref{fig:fcc_kappa}(a)-\ref{fig:fcc_kappa}(c) show the cumulative time-average of the thermal conductivity as computed using Eq.~(\ref{equation:kappa}) for the \gls{bpf} structure at 300 K. A cubic simulation cell with 30,000 atoms was used. Apart from the \gls{nep}-C$_{60}$ potential model constructed in this work, we also considered the old \gls{nep}-Carbon model \cite{fan2022arxiv} and a hybridized potential formed by an intra-molecular Tersoff potential \cite{lindsay2010prb} and an inter-molecular \gls{lj} potential, which we call the Tersoff-\gls{lj} potential. The energy and length parameters in the \gls{lj} potential was chosen as $\epsilon=$ 2.86 meV and $\sigma=$ 3.47 \AA{} \cite{girifalco2000prb}. 
The time-converged thermal conductivity is presented in Fig.~\ref{fig:fcc_kappa}(d) and Table~\ref{table:kappa}. 

\begin{table}[htb]
\centering
\setlength{\tabcolsep}{1.8Mm}
\caption{Thermal conductivity $\kappa$ (in units of W/mK) of \gls{bpf} and \gls{qhpf} (in the $x$ and $y$ directions) as calculated by NEP-Carbon, NEP-C$_{60}$, Tersoff and from experiments. An \gls{lj} potential is added to the Tersoff potential in the case of \gls{bpf}. The experimental value is taken from Ref. \cite{Yu1992prl}}
\label{table:kappa}
\begin{tabular}{lllll}
\hline
\hline
\multicolumn{2}{c}{Method }            & \multicolumn{1}{c}{\gls{qhpf}-$x$} & \multicolumn{1}{c}{\gls{qhpf}-$y$} & \multicolumn{1}{c}{\gls{bpf}} \\
\hline
\multirow{3}{*}{HNEMD} & Tersoff       & $75(9)$     & $173(7)$      & $0.14(6)$ \\
                       & NEP-Carbon    & $34(1)$     & $37(2)$       & $1.6(3)$  \\
                       & NEP-C$_{60}$  & $102(3)$    & $137(7)$     & $0.45(5)$  \\
EMD                    & NEP-C$_{60}$  & $109(19)$    & $138(18)$  & NA    \\
\multicolumn{2}{l}{Experiment}         & NA     & NA         & $\sim 0.4$   \\
\hline
\hline
\end{tabular}
\end{table}

The thermal conductivity of \gls{bpf} at 300 K has been measured to be about $0.4$ W/mK \cite{Yu1992prl}. The Tersoff-\gls{lj} potential gives $0.14(6)$ W/mK, which is much smaller than the experimental value. The \gls{nep}-Carbon model gives $1.6(3)$ W/mK, which is much larger than the experimental value. The \gls{nep}-C$_{60}$ model, on the other hand, gives $0.45(5)$ W/mK, which is very close to the experimental value. Moreover, the fast rotation of the C$_{60}$ molecules in \gls{bpf} as observed experimentally \cite{Yu1992prl} can been reproduced with the \gls{nep}-C$_{60}$ model (see the Supplementary Material for movies of the trajectories during the \gls{md} simulations of \gls{bpf} as well as \gls{qhpf}) but not with the \gls{nep}-Carbon model. After confirming the reliability of the \gls{nep}-C$_{60}$ model in the prediction of the thermal conductivity of \gls{bpf}, we next study heat transport in the new \gls{qhpf} structure.

\subsubsection{The monolayer quasi-hexagonal-phase fullerene}

\begin{figure}
\centering
\includegraphics[width=\columnwidth]{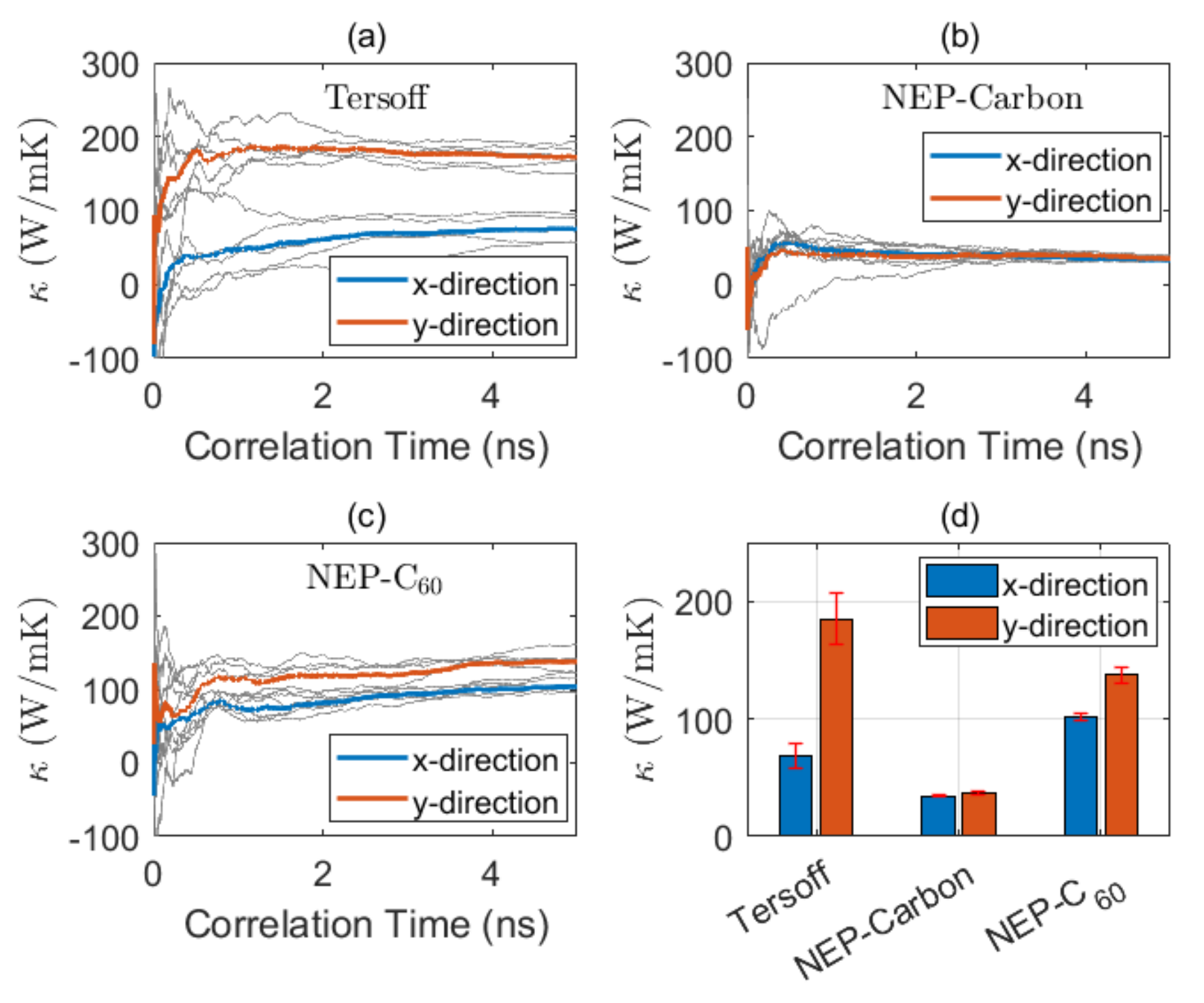}
\caption{
Thermal conductivity as a function of time for the \gls{qhpf} structure at 300 K calculated using the \gls{hnemd} method with (a) the Tersoff potential, (b) the \gls{nep}-Carbon model, and (c) the \gls{nep}-C$_{60}$ model. (d) The converged thermal conductivity values from the different potentials.
}
\label{fig:qHPF_kappa}
\end{figure}

\begin{figure}
\centering
\includegraphics[width=\columnwidth]{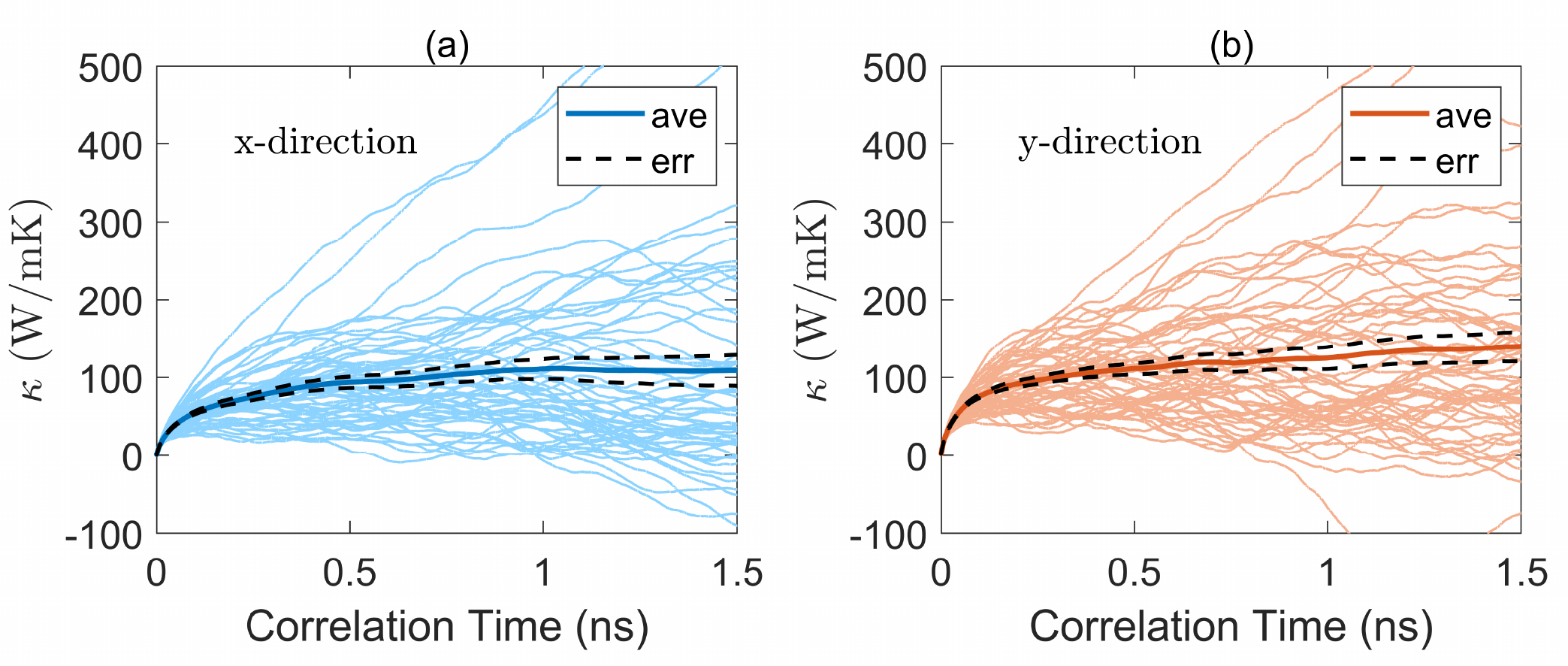}
\caption{
The running thermal conductivity as a function of the correlation time for the \gls{qhpf} structure at 300 K in the (a) $x$ and (b) $y$ directions calculated using the \gls{emd} method with the \gls{nep}-C$_{60}$ potential. In each subplot, the thin lines represent the results from 60 individual runs (each with 5 ns production time), and the thick solid and dashed lines represent their average and error bounds.
}
\label{fig:emd}
\end{figure}

We similarly calculated the thermal conductivity of the \gls{qhpf} structure at 300 K using the three potential models. A rectangular cell with 28,800 atoms was used, which is large enough to eliminate finite-size effects in the \gls{hnemd} method. The results are shown in Fig.~\ref{fig:qHPF_kappa} and Table \ref{table:kappa}. As a cross-check, we also calculated the thermal conductivity of the \gls{qhpf} structure at 300 K using the \gls{emd} method (with the same simulation domain size as in \gls{hnemd}). The results are shown in Fig.~\ref{fig:emd} and Table \ref{table:kappa}. We see that the results from \gls{emd} and \gls{hnemd} are consistent in both the $x$ and $y$ directions within the statistical error bounds.

Different from \gls{bpf}, which is essentially isotropic regarding heat transport, it turns out that heat transport in \gls{qhpf} is anisotropic. The thermal conductivity in the $y$ direction (see Fig.~\ref{fig:model}) is about 40\% higher than that in the $x$ direction according to the \gls{nep}-C$_{60}$ potential. The $x$ and $y$ components of the thermal conductivity in \gls{qhpf} are about 200 and 300 times of that of \gls{bpf}. The \gls{nep}-Carbon potential predicted a much smaller anisotropy while the Tersoff potential predicted a much larger one. Based on the results for \gls{bpf}, we expect that the \gls{nep}-C$_{60}$ potential gives the most reliable predictions due to its superior accuracy for the C$_{60}$-based structures. Nevertheless, we stress that the thermal conductivity of \gls{qhpf} has not been experimentally measured so far and out results here should be regarded as theoretical predictions.

\begin{figure}
\centering
\includegraphics[width=\columnwidth]{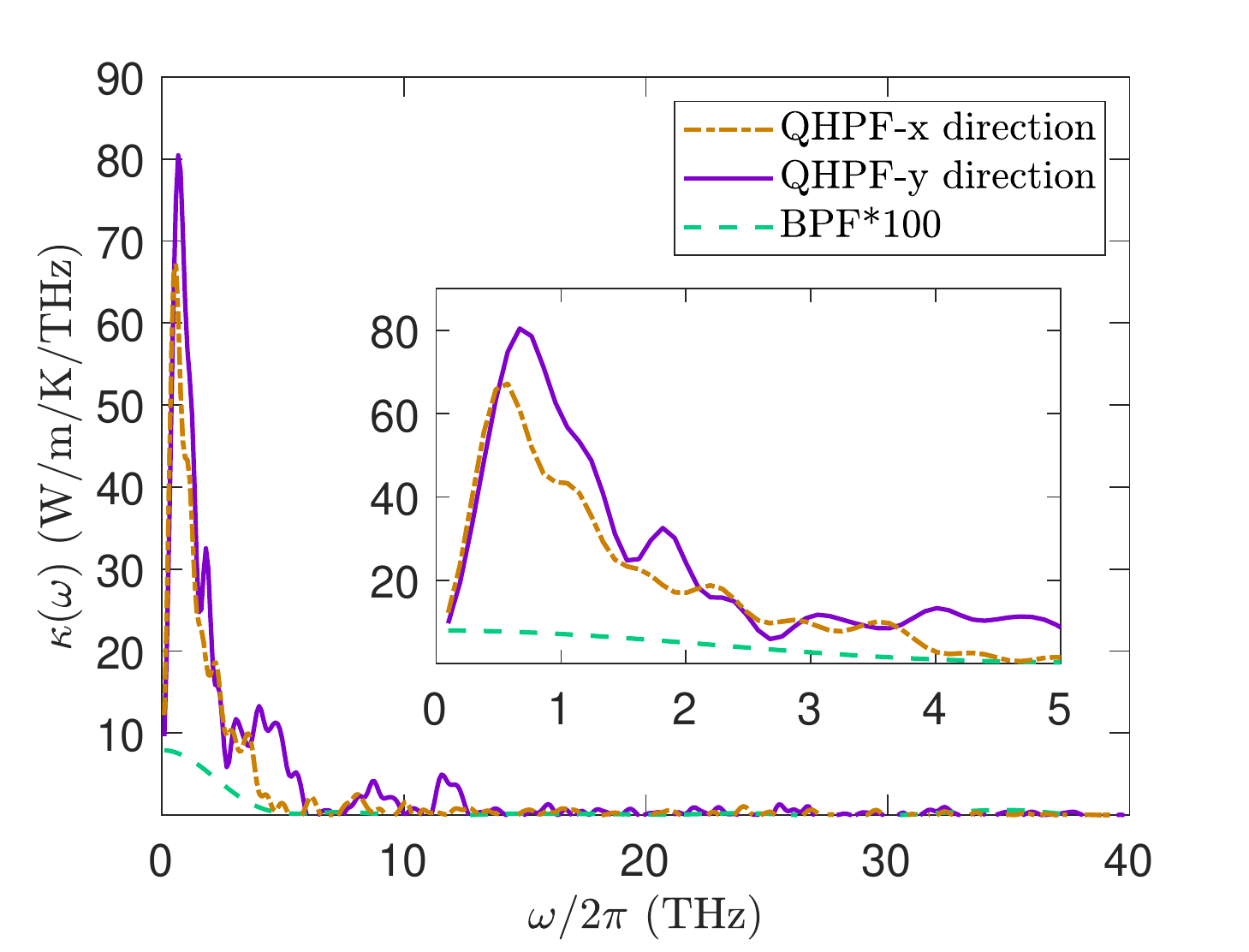}
\caption{
The spectral thermal conductivity as a function of the vibrational frequency for \gls{qhpf} in the $x$ and $y$ directions and \gls{bpf} calculated using the \gls{nep}-C$_{60}$ potential. Note that we have multiplied the $\kappa(\omega)$ for \gls{bpf} by 100. The inset shows the part with $\omega/2\pi<5$ THz.
}
\label{fig:shc}
\end{figure}

\begin{figure}
\centering
\includegraphics[width=\columnwidth]{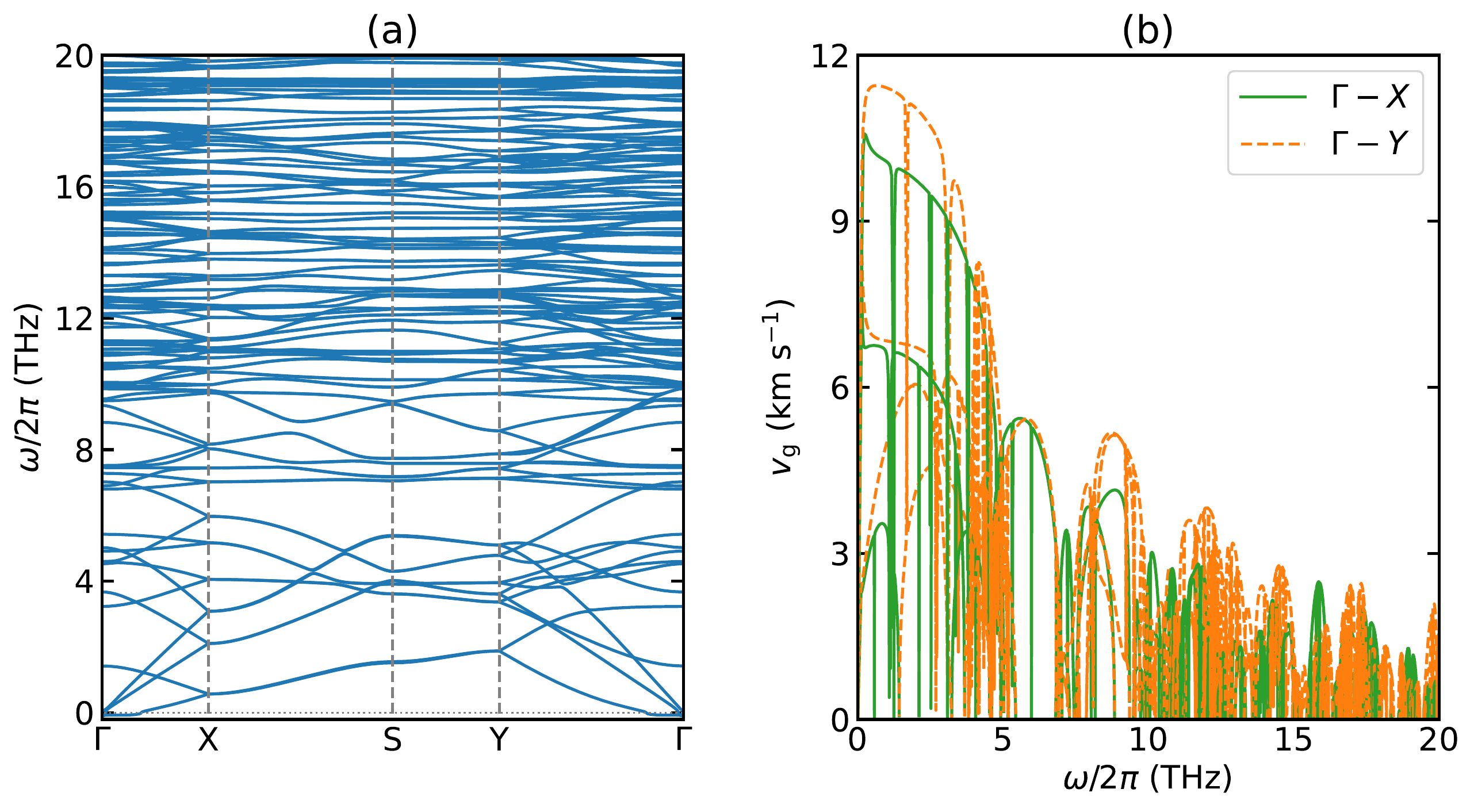}
\caption{
(a) Phonon dispersion relations and (b) phonon group velocities in \gls{qhpf}. We have omitted frequencies above 20 THz for clarity. The small negative frequencies around the $\Gamma$ point in (a) are typical for the frozen-phonon approach we used. More advanced methods that take finite-temperature effects into account can help to alleviate this numerical problem. 
}
\label{fig:dispersion}
\end{figure}

To gain more insight, we show the spectral thermal conductivity $\kappa(\omega)$ of \gls{qhpf} in the $x$ and $y$ directions and that of \gls{bpf} in Fig.~\ref{fig:shc}. We see that for all the materials and transport directions considered here, heat is mainly transported by phonons with frequency smaller than about $\nu=\omega/2\pi=2$ THz, which is almost the range for the acoustic phonon branches in \gls{qhpf} (see Fig.~\ref{fig:dispersion}(a)). This means that the acoustic phonons are the major heat carriers in \gls{qhpf}. On the other hand, the phonon group velocities (see Fig.~\ref{fig:dispersion}(b)) show relatively larger values for the $y$ direction (corresponding to the $\Gamma-Y$ path) than the $x$ direction (corresponds to the $\Gamma-X$ path). This can partially explain the anisotropic thermal conductivity in \gls{qhpf} as group velocity is one of the major factors determining the thermal conductivity. Besides thermal conductivity, the in-plane elasticity in \gls{qhpf} has also been found to be anisotropic \cite{Ying2023eml}. The anisotropy as exhibited by these physical properties can be expected from the asymmetric inter-molecular bonding in \gls{qhpf} as shown in Fig.~\ref{fig:model}.

\begin{figure}
\centering
\includegraphics[width=\columnwidth]{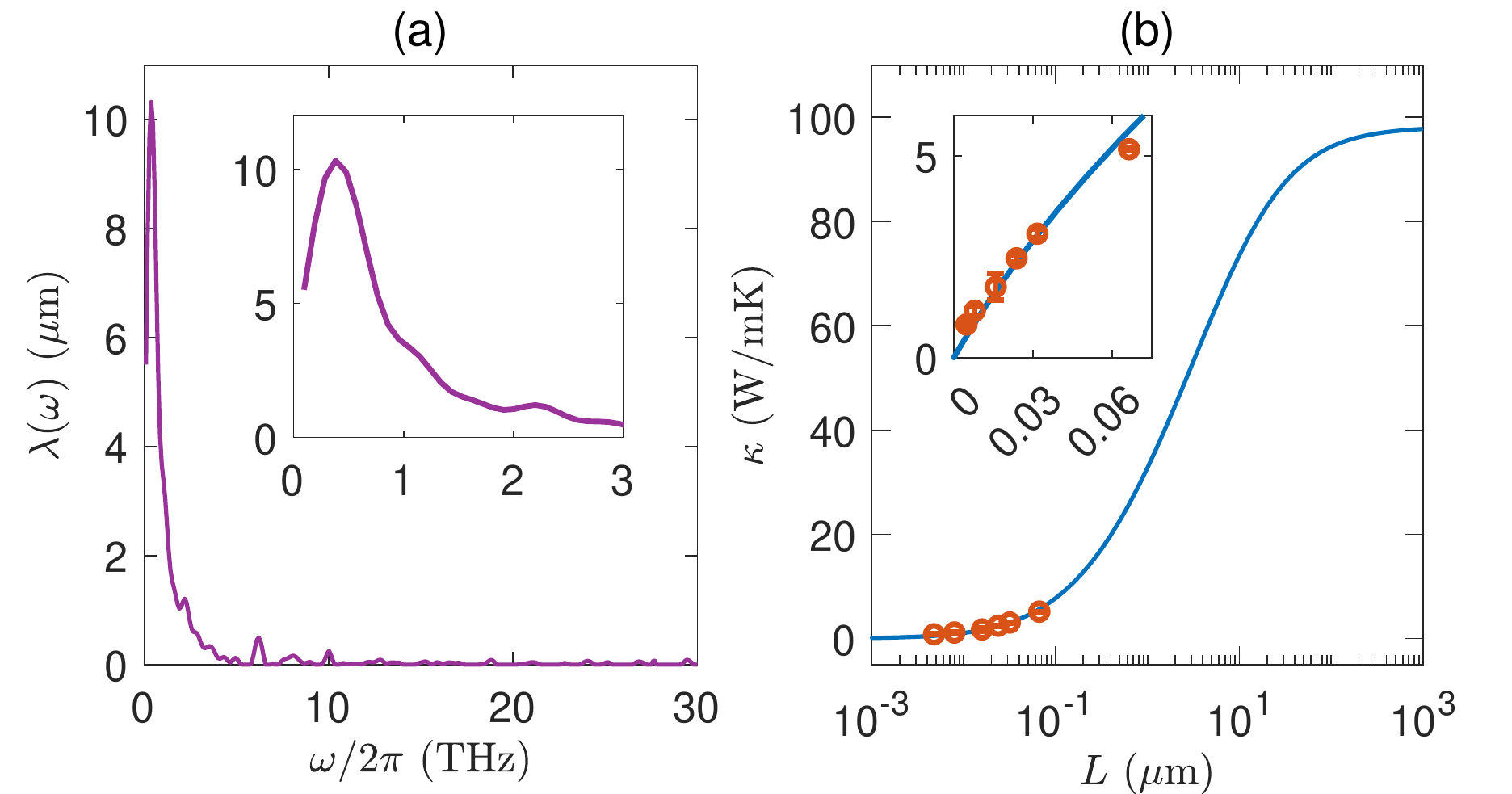}
\caption{(a) The phonon mean free path $\lambda(\omega)$ as a function of the frequency for \gls{qhpf} in the $x$ direction calculated using the \gls{nep}-C$_{60}$ potential. The inset shows the part with $\omega/2\pi<3$ THz. (b) The effective thermal conductivity $\kappa$ as a function of the system length $L$. The  circles represent results from the NEMD simulations, and the line is from the \gls{hnemd} simulations. The inset shows the part with $L<80$ nm.
}
\label{fig:mfp}
\end{figure}

Using the methods as developed in Refs.~\cite{fan2019prb} and \cite{li2019jcp}, we also calculated the phonon \gls{mfp} spectrum $\kappa(\omega)$, as shown in Fig.~\ref{fig:mfp}(a). The low-frequency phonons develop \glspl{mfp} up to about 10 microns, and the thermal conductivity thus only exhibits a convergence up to a system length (not to be confused with the simulation domain length in the \gls{hnemd} or \gls{emd} methods) of about one millimeter, as can be seen from  Fig.~\ref{fig:mfp}(b). Due to the presence of the large \glspl{mfp}, it is unfeasible to calculate the diffusive thermal conductivity using the \gls{nemd} method. However, agreement between \gls{nemd} and \gls{hnemd} can be observed in a range of system length  that is affordable for our \gls{nemd} simulations.

\begin{figure}
\centering
\includegraphics[width=\columnwidth]{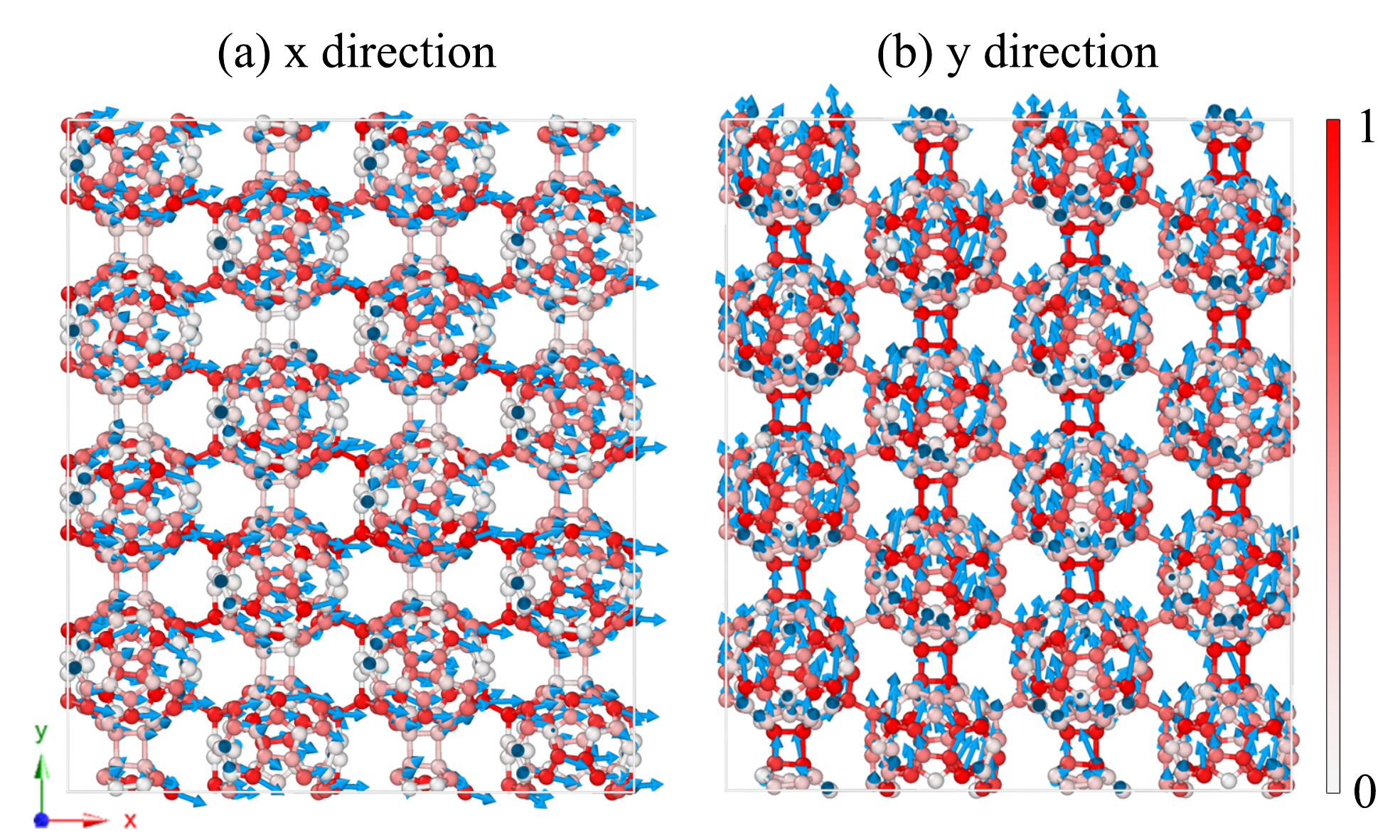}
\caption{
    The per-atom heat current distribution in \gls{qhpf} with heat transport in the (a) $x$ and (b) $y$ directions. The color on the atoms represents the normalized magnitude of the per-atom heat current in the transport direction. The arrow represents both the magnitude and the direction of the per-atom heat current. Only a small part of a system with 28,800 atoms is shown here for clarity.
}
\label{fig:heat_current}
\end{figure}

Apart from the frequency and momentum spaces, insight can also be gained by from the real space. Figure~\ref{fig:heat_current} shows the real-space heat current distribution in the \gls{hnemd} simulations. When the transport direction is $x$, inter-molecular heat is mainly carried by the atoms forming the C-C single bonds between the molecules; when the transport direction is $y$, inter-molecular heat is mainly carried by the atoms forming the so-called $[2+2]$ cycloaddition bonds along the [010] direction \cite{hou2022nature}. For both directions, the inter-molecular heat is not mainly carried by the weak \gls{vdw} interactions but by the strong covalent bonds. By contrast, there is no persistent inter-molecular covalent bond in \gls{bpf} and the C$_{60}$ molecules rotate quickly, resulting in a strong suppression of the heat transport. This comparison highlights the vital role played by the inter-molecular covalent bonds in enhancing the thermal conductivity in \gls{qhpf} as compared to that in \gls{bpf}.

\section{Summary and conclusions}

In summary, we have constructed an accurate and transferable \gls{mlp}  based on the efficient \gls{nep} approach \cite{fan2021neuroevolution}, which is applicable to both \gls{bpf} and \gls{qhpf}. The \gls{nep} model can accurately describe both the covalent bonding and the \gls{vdw} interactions in the C$_{60}$ based structures. It predicted a thermal conductivity value of $0.45(5)$ W/mK at 300 K for \gls{bpf}, which agrees with the experimental results \cite{Yu1992prl} excellently. We then predicted the thermal conductivity of \gls{qhpf} to be anisotropic and is more than two orders of magnitude higher than \gls{bpf}. We find that the inter-molecular covalent bonding in \gls{qhpf} plays a crucial role in enhancing the thermal conductivity in \gls{qhpf} as compared to that in \gls{bpf}. As a possible future direction of research, we note that the \gls{nep}-C$_{60}$ model developed here can be extended (by adding extra training data) to study multi-layer and three-dimensional structures of stacked \gls{qhpf} as well as other polymerized structures based on C$_{60}$.

\section*{Data availability}
Complete input and output files for the \gls{nep} training and testing are freely available at a Zenodo repository \cite{penghua_ying_2022_6972675}. 

\begin{acknowledgments}
H.D. and Z.F. acknowledge support from the National Natural Science Foundation of China (NSFC) (No. 11974059) and the Research Fund of Bohai University  (No. 0522xn076).
C.C. and P.Q. acknowledge the support from the National Key Research and Development Program of China (2021YFB3802104).
P.Y. thanks Jin Zhang, Ting Liang, Xiaowen Li, and Xiaobin Qiang for valuable discussions.
We acknowledge the computational resources provided by High Performance Computing Platform of Beijing Advanced Innovation Center for Materials Genome Engineering.
\end{acknowledgments}

\bibliography{refs}

\end{document}